\documentclass[a4paper,10pt]{article}
\usepackage[utf8]{inputenc}
\usepackage{amssymb,amsmath}
\usepackage{authblk}
\usepackage[left=0.5in, right=0.5in, top=1in, bottom=1in]{geometry}
\usepackage{hyperref}
\usepackage{graphicx}
\usepackage{rotating}
\usepackage{multirow}
\usepackage{array,booktabs}
\usepackage{caption}

\title{Virial coefficients from unified statistical 
thermodynamics of quantum gases trapped under generic power law potential in $d$ dimension 
and the equivalence of   quantum gases}

\author{  Shah Mohammad Bahauddin$^{a}$, 
Mir Mehedi Faruk*$^{b,c}$\\
 Physics and Astronomy, Rice University, Houston, USA$^a$.
Theoretical Physics, Blackett Laboratory, Imperial College, London SW7 2AZ, United Kingdom$^b$\\
Department of Theoretical Physics, University of Dhaka, Dhaka-1000$^c$\\

\href{mailto:me@somewhere.com}{*Email: muturza3.1416@gmail.com, mmf15@ic.ac.uk} 
 }

\begin{document}
\large
\maketitle
 
 \begin{abstract}
 From the
 unified statistical thermodynamics of quantum gases, 
 the virial coefficients of ideal Bose and Fermi gases, 
 trapped under 
 generic power law potential 
 are derived systematically.
 From the general result of virial coefficients,
 one can produce the known results in $d=3$ and $d=2$. 
But more importantly
 we found that,
 the virial coefficients of Bose and Fermi gases  become identical 
 (except the second virial coefficient, where the sign is different)
 when  the gases are trapped under harmonic
 potential in $d=1$. This result suggests the equivalence between Bose and Fermi gases 
 established in $d=1$ (J Stat Phys, DOI 10.1007/s10955-015-1344-4). 
 Also, it is found that the virial coefficients of two-dimensional free Bose (Fermi) gas are equal to 
 the virial coefficients of one-dimensional harmonically trapped Bose (Fermi) gas.
 \end{abstract}
\section{Introduction}
After the demonstration by Einstein\cite{bose,einstein} that there
is a possibility of condensation of free bosons, bulk 
behavior of ideal free Bose gas are studied by many authors\cite{ziff,may,r,lieb1,lieb2}.
At the same time, theoretical investigations are also done for the free Fermi gas \cite{b1,b2,b3,b4,b5}
and remarkable property of degeneracy pressure of the free fermions
are noticed\cite{pathria}. 
Dimensional dependence of the 
thermodynamic quantities
of both types
of ideal quantum gases are investigated in detail\cite{ziff,fisher,beckmann}
and interesting dimensional
dependence of thermodynamic quantities was reported.
For instance, BEC is noticed in free Bose gas only when $d>2$, whether the specific heat is found discontinuous if $d\geq 4$. 
But, it was May\cite{may},
who first noticed an equivalence of specific heat for two-dimensional ideal free 
bosons and fermions.
Later, Lee\cite{lee} generalized this equivalence between two-dimensional
quantum gases for other thermodynamic quantities as well.
It is reported in the paper that, one can obtain this equivalence,
if the fugacities of Bose and Fermi gases are related by Euler transformation\cite{lee,lee1}.
He also found a way to present the thermodynamic quantities of both of Fermi and Bose gases
in a unified approach.
Vieferes et. al. \cite{Vieferes} showed
that all the virial coefficients are equal in two
dimensional free quantum gases (except the second virial coefficients 
for which the signs are opposite).
\\\\
The subject of quantum gases
drew more
attention after it was possible to experimentally 
detect BEC\cite{Bradley,anderson,davis} and Fermi degeneracy\cite{DeMarco}
in trapped quantum
gases. Since then,
a lot of 
studies are done 
on Bose and Fermi gas trapped under generic power\cite{s,sala,yan,yan1,yan2,yan3,m1,champu}
law potential ($U= \sum_{i=1} ^d c_i |\frac{x_i}{a_i}|^{n_i}$) in arbitrary dimension. Some drastic changes are noted
in the characteristics
of both Bose and Fermi 
gases due to trapping potential\cite{s,sala,yan,yan1,yan2,yan3,m1,champu}.
For instance, Bagnato and Kleppner demonstrated the possibility of BEC of an ideal Bose gas confined by one- and two-dimensional power-law traps \cite{Bagnato,Dai} ($d<3$)
with appropriate trapping potential, which is not a property of free Bose gas\cite{sala,m1}.
In addition, the equivalence of two-dimensional free
quantum gases do not remain valid for any trapping potential\cite{turza}.
Very recently Mehedi\cite{turza} was able to present the thermodynamics of both types of
quantum gases trapped under generic power law potential
in a unified way. 
Remarkably an equivalence between them in d=1 was found for harmonically trapped systems\cite{turza}.
Turning our attention towards virial coefficients, it would be intriguing
to check if the virial coefficients
become equal, for harmonically trapped quantum gases in one dimension.\\\\
Virial coefficients appear to relate the pressure of a many-particle system in powers of the number density 
in a given $d$-dimensional volume, providing a systematic treatment in the corrections of the laws of ideal gases.
The $d$ 
dimensional virial coefficients of quantum gases 
trapped under generic power-law potentials are not yet reported.
In this paper, we have first calculated the virial coefficients of both types
of ideal quantum gases 
trapped under generic power law potential ($U= \sum_{i=1} ^d c_i |\frac{x_i}{a_i}|^{n_i}$)
in a unified approach in arbitrary
dimension.
But the main motivation of this investigation
is to check out whether 
the virial coefficients of Bose and Fermi gases become
same for when they are harmonically trapped in $d=1$.
  Now, from the more general result one should be able to reproduce the known solutions
such as text book results of virial coefficients of bosons and fermions in $d$=3 and the outcome of Vieferes et. al.'s calculation relating virial coefficients of two-dimensional free Bose and Fermi gases. 
But more importantly, we would be able to find out,
whether the virial coefficients are also the same 
for one-dimensional quantum gases
trapped in a harmonic potential,
suggesting the equivalence found by Mehedi\cite{turza}.\\\\
\section{Virial coefficients of ideal quantum gases trapped under generic power law potential}
For a quantum gas, the average number of particles occupying the $i$-the single particle energy 
eigenstate and the grand potential  are
 given by
\begin{eqnarray}
\bar{n}_i &=&\frac{1}{z^{-1}e^{\beta\epsilon_i}+a}
\end{eqnarray}and,
\begin{eqnarray}
q&=&\frac{1}{a}\sum_\epsilon ln(1+az e^{-\beta \epsilon})
\end{eqnarray}
where, $a=-1(1)$ stands for a Bose systems (Fermi systems),
$z$ is the fugacity and $\beta=\frac{1}{KT}$ is the Boltzmann constant.
Let us consider an ideal quantum system trapped in a generic power law potential
in $d$ dimensional space with  a single particle Hamiltonian of the form,
\begin{eqnarray}
 \epsilon (p,x_i)= bp^l + \sum_{i=1} ^d c_i |\frac{x_i}{a_i}|^{n_i}
\end{eqnarray}\\
where, $b,$ $l,$ $a_i$, $c_i$ and $n_i$  
are all positive constants, $p$ is the momentum 
and $x_i$ is the  $i$ th component of the coordinates of a particle.
Here, $c_i$, $a_i$, $n_i$ determine the depth 
and confinement power of
the potential, $l$ is the kinematic parameter, and $x_i<a_i$. As $|\frac{x_i}{a_i}|<1$,
the potential term goes to zero
as all $n_i\longrightarrow \infty$.
 Using $l=2$, $b=\frac{1}{2m}$ one can get the 
 energy spectrum  of  the hamiltonian used literature \cite{pathria,ziff,sala}.
If one uses $l=1$ and $b=c$ one finds the hamiltonian of massless systems such as photons\cite{pathria}.
Now replacing the sum by integration,
the grand potential for the quantum gases become\cite{turza},
\begin{eqnarray}
 q=sign(\sigma)\frac{V_d'}{\lambda'^d} Li_{\chi+1} (\sigma)
\end{eqnarray}
where, $\lambda'$ and $V'$ is the effective thermal wavelength and effective volume,
\footnote{To read more about 
effective volume and effective thermal wavelength,
see \cite{yan,yan1, yan3}} and $Li_{\chi} (\sigma)$ is the polylog
function. Here,
\begin{eqnarray}
       V_d ' &=& V_d \prod_{i=1} ^d \Bigg(\frac{kT}{c_i} \Bigg)^{1/n_i}\Gamma \Bigg(\frac{1}{n_i} + 1 \Bigg),\\
     \lambda'&=& \frac{h b^{\frac{1}{l}} \beta ^{\frac{1}{l}}}{\pi ^{\frac{1}{2}}} \Bigg[\frac{\Gamma (d/2+1)}{\Gamma (d/l+1)} \Bigg]^{1/d}.
\end{eqnarray}
And a useful representation of polylog is\cite{lee}
\begin{equation}
 Li_{q}(m)=\frac{1}{\Gamma(q)}\int_0 ^m \Bigg[\ln \bigg(\frac{m}{\eta} \bigg) \Bigg]^{q-1}\frac{d\eta}{1-\eta}, 
\end{equation}
for $Re(m)<1$. And quantity $\chi$ and $\sigma$ are,
\begin{eqnarray}
\chi=\frac{d}{l}+\sum_{i=1} ^{d}\frac{1}{n_i},\\
\sigma = \left\{
     \begin{array}{lr}
     -z &,  \text{Fermi system}\\
     z &,  \text{Bose system}.
     \end{array}
   \right.
\end{eqnarray}\\
where $z$ is the fugacity. 
Now the density can be calculated from the grand potential,
\begin{eqnarray}
&& N=z \bigg(\frac{\partial Q}{\partial z} \bigg)_{\beta,V}\nonumber \\
&&\Rightarrow \rho=\frac{N}{V_d '}=sgn(\sigma)\frac{1}{\lambda'^d} Li_\chi(\sigma)
\end{eqnarray}
And the pressure,
\begin{eqnarray}
 P=\frac{1}{\beta} \bigg(\frac{\partial Q}{\partial V_d '} \bigg)_{\beta,z}\nonumber,\\
\Rightarrow \beta P=\frac{1}{\lambda'^d} Li_{\chi+1}(\sigma).
\end{eqnarray}
Now writing the the pressure and density equation as series\cite{pathria}
\begin{eqnarray}
 \beta P = \frac{1}{\lambda'^d} \bigg(\sum_{j=1} ^\infty b_j z^j \bigg),\\
 \rho =\frac{1}{\lambda '^d} \bigg(\sum_{j=1} ^\infty j b_j z^j \bigg).
\end{eqnarray}\\
Where the coefficients $b_j$ 
are defined as,
\begin{eqnarray}
&& b_j = \left\{    \begin{array}{lr}
     (-1)^{j+1} j^{-(\chi+1)} &,  \text{Fermi system}\\
     j^{-(\chi+1)} &,  \text{Bose system}.
     \end{array}
   \right.
\end{eqnarray}\\
Next we need to express the fugacity in
 terms of density so that one can write the pressure as a series of density. Writing the density equation explicitly,
\begin{eqnarray}
 \lambda '^d\rho=b_1z +2b_2 z^2+3b_3z^3+4b_4z^4+...
\end{eqnarray}
In the high temperature limit we can approximate fugacity,
\begin{equation}
 z=\sum_{j=1} ^{\infty}a_j (\rho \lambda'^d)^j=a_1( \lambda '^d\rho)+a_2( \lambda '^d\rho)^2+a_3( \lambda '^d\rho)^3+...
\end{equation}\\
Putting $z$
into the expression $\lambda '^d\rho$,
\begin{eqnarray}
 \lambda '^d\rho=
 &&b_1 (a_1( \lambda '^d\rho)+a_2( \lambda '^d\rho)^2+a_3( \lambda '^d\rho)^3+...)
 +2b_2 (a_1( \lambda '^d\rho)+a_2( \lambda '^d\rho)^2 + a_3( \lambda '^d\rho)^3+...)^2 +\nonumber\\ 
&& 3b_3(a_1( \lambda '^d\rho)+a_2( \lambda '^d\rho)^2 + a_3( \lambda '^d\rho)^3+...)^3
 \nonumber \\
=&& \lambda '^d\rho (b_1a_1) + (\lambda '^d\rho)^2(b_1a_2+2b_2 a_1^2)+ (\lambda '^d\rho)^3 (b_1 a_3+4b_2a_1a_2+3b_3a_1^3) +  (\lambda '^d\rho)^4 (b_1 a_4 + \nonumber\\ 
&& 2b_2a_2^2 + 4b_2a_1a_3+9b_3a_1^2a_2+4b_4a_1^4) + (\lambda '^d\rho)^5 (b_1 a_5 + 4b_2a_2a_3 + 4b_2a_1a_4 + 9b_3a_1a_2^2 + \nonumber\\ 
&& 9b_3a_1^2a_3 + 16b_4a_1^3a_2+5b_5a_1^5)...
\end{eqnarray}\\
Now comparing the equation (15) and (17),
\begin{eqnarray}
&& b_1 a_1=1,\nonumber\\
 &&b_1a_2+2b_2 a_1^2=0,\\
&& b_1a_3+4b_2 a_1 a_2 +3b_3 a_1 ^3=0,\nonumber\\
 &&................................\nonumber\\
&&................................\nonumber
 \end{eqnarray}\\
So, one can establish relation between $a_i$ and $b_i$,
\begin{eqnarray}
&& b_1=\frac{1}{a_1} \Leftrightarrow  a_1=\frac{1}{b_1}, \nonumber\\
&& b_2=-\frac{a_2}{2} \Leftrightarrow a_2=-2b_2,\nonumber\\
&& b_3= \frac{1}{3}(2a_2^2-a_3) \Leftrightarrow a_3=8b_2 ^2 -3b_3,\\
&&................................\nonumber\\
&&................................\nonumber
\end{eqnarray}\\
Now from the pressure equation,
\begin{eqnarray}
 \beta P&=&\frac{1}{\lambda '^d} \bigg(\sum_{j=1} ^\infty b_j z^j \bigg)\nonumber\\
 &=& \frac{1}{\lambda '^d}(b_1 z+ b_2 z^2 +b_3 z^3+...)\nonumber \\
 &=& \frac{1}{\lambda '^d} (b_1 (a_1 (\rho \lambda'^d) +a_2 (\rho \lambda'^d)^2 +a_3 (\rho \lambda'^d)^3+...)
 + b_2  (a_1 (\rho \lambda'^d) +a_2 (\rho \lambda'^d)^2 +a_3 (\rho \lambda'^d)^3+...)^2\nonumber \\
&& +b_3 (a_1 (\rho \lambda'^d) +a_2 (\rho \lambda'^d)^2 +a_3 (\rho \lambda'^d)^3+...)^3+... ... ...)\nonumber\\
 &=&\frac{1}{\lambda '^d} (  (\rho \lambda'^d) +  
 (-b_2)(\rho \lambda'^d)^2 +
 (4b_2^2 -2b_3)(\rho \lambda'^d)^3 +
(-20b_2^3+18b_2b_3-3b_4) (\rho \lambda'^d)^4
+(112 b_2 ^4\nonumber\\
&& -144 b_2^2 b_3 + 18 b_3^2 +32 b_2b_4 -4b_5) (\rho \lambda'^d)^5 + 
(-672b_2 ^5+1120 b_2 ^3 b_3-315b_2b_3^2-280b_2^2b_4\nonumber\\
&& +60b_3b_4+50b_2b_5-5b_6) (\rho \lambda'^d)^6+(4224b_2^6
-8640b_2^4b_3
+3888b_2 ^2 b_3 ^2-216b_3^3 +2304 b_2^3b_4\nonumber\\
&& -1152b_2b_3b_4
+48b_4^2-480b_2^2 b_5+90b_3b_5+72b_2b_6-6b_7)(\rho \lambda'^d)^7+..)
 \end{eqnarray}\\
The virial expansion is defined as,
\begin{eqnarray}
 \beta P=\sum_{k=1} ^{\infty}A_k \rho ^k \lambda ^{d(k-1)}.
\end{eqnarray}\\
where, $A_l$ are the virial coefficients.
Thus comparing the above two equations we can calculate the virial coefficients,
\begin{eqnarray}
&& A_1=1,\\
&& A_2=-b_2,\\
&& A_3= 4b_2^2 -2b_3,\\
&&A_4= -20b_2^3+18b_2b_3-3b_4,\\
&&A_5= 112 b_2 ^4-144 b_2^2 b_3 + 18 b_3^2 +32 b_2b_4 -4b_5,\\
&&A_6=-672b_2 ^5+1120 b_2 ^3 b_3-315b_2b_3^2-280b_2^2b_4+ 60b_3b_4+50b_2b_5-5b_6,\\
&&A_7=4224b_2^6 -8640b_2^4b_3 +3888b_2 ^2 b_3 ^2 -216b_3^3 +2304 b_2^3b_4 -1152b_2b_3b_4+48b_4^2\nonumber\\
&&\quad \quad-480b_2^2 b_5+90b_3b_5+72b_2b_6-6b_7,\\
&&.............................................\nonumber\\
&&..............................................\nonumber
\end{eqnarray}
where, the  $b_j$ are defined in equation (14). Point to note that, these results are quite general for ideal quantum gas trapped under generic power law potential.
Depending upon boson or fermion as well as the type of trapping potential the virial coefficients, eq. (22) - (28) will change accordingly. Also, the result of eq. (22) - (25) coincides well with \cite{igu}.
\section{Results}
 The results of virial coefficients presented in eq. (22)-(28)
are for ideal 
quntum gases trapped under generic power law potential in any dimension. In this section we first
present the final results for some 
 specific  situation.
We have later presented two tables containing the results
of virial coefficients of bosons and fermions, trapped under different power law potentials.
\subsection{Free quantum gases in d=3}
At first, we present the  virial coefficients
for 
the free massive bosons ($l=2$) in three dimensional space. So,
choosing
$d=3$ and all $n_i\longrightarrow \infty$ which leaves us $\chi=\frac{3}{2}$. 
In this case eq. (22)-(28) dictates, the virial coefficients of bosons are,
\begin{eqnarray}
&& A_1=1,\nonumber\\
&& A_2=-0.177,\nonumber\\
&& A_3=-0.033,\nonumber\\
&& A_4=-0.00011,\nonumber\\
&& A_5=-0.00000354,\nonumber\\
&&.............................................\\
&&..............................................\nonumber
\end{eqnarray}
Turning our attention towards free massive Fermi gas in three dimensional space,
the virial coefficients are,
\begin{eqnarray}
 && A_1=1,\nonumber\\
&& A_2=0.177,\nonumber\\
&& A_3=-0.033,\nonumber\\
&& A_4=0.00011,\nonumber\\
&& A_5=-0.00000354,\nonumber\\
&&.............................................\\
&&..............................................\nonumber
\end{eqnarray}
The results of eq. (29) and (30) concurs with Ref. \cite{pathria}.
It is interesting to point out that, in the case of Fermi gases in $d=3$ the virial coefficients alters sign 
while all the virial coefficients of Bose gases are negative.
Point to note, any negative (positive)
virial coefficient $A_n$ indicates that, the $n$-particle wave function is symmetric (asymmetric). 
As the bosonic
$n$- particle wave function is 
always symmetric in three-dimensional space, we see the virial coefficients of bosons are
negative. 
And as we know in three dimensions,
the $n$ particle fermion 
wavefunctions (even $n$) are anti-symmetric, $A_n$ are positive for even $n$, in $d=3$.
But any $n$-particle Fermi wavefunction is symmetric in d=3 while $n$ is odd. So,
we have found  negative virial coefficient $A_n$, for odd n.
Thus we find out the reason due to which we get the alternating sign in virial coefficients of 
Fermi gas in $d=3$.
\\
\subsection{Free quantum gases in d=2}
Turning our focus on the virial coefficients of free massive quantum gases in lower dimensions
we choose $d=2$ and all $n_i \longrightarrow \infty$, and find out that $\chi=2$.
Eq. (22)-(28) dictates the virial coefficients for bosons are,
\begin{eqnarray}
&& A_1=1,\nonumber\\
&& A_2=-0.25,\nonumber\\
&& A_3=0.0278,\nonumber\\
&& A_4=0,\nonumber\\
&& A_5=-0.000278,\nonumber\\
&& A_6=0,\nonumber \\
&&A_7=0.00000472,\nonumber \\
&&.............................................\\
&&..............................................\nonumber
\end{eqnarray}
while the virial coefficients of fermions are,
\begin{eqnarray}
&& A_1=1,\nonumber\\
&& A_2=0.25,\nonumber\\
&& A_3=0.0278,\nonumber\\
&& A_4=0,\nonumber\\
&& A_5=-0.000278,\nonumber\\
&& A_6=0,\nonumber \\
&&A_7=0.00000472,\nonumber \\
&&.............................................\\
&&..............................................\nonumber
\end{eqnarray}
The results matches precisely with the viefers et. al.\cite{Vieferes}.
Please note that, all the
virial coefficients for Bose gases are equal to Fermi gases (except for the second virial coefficient,
for which the signs are opposite).
This important result turns out to indicate the equivalence of free
Bose and Fermi gases in two dimension\cite{lee}.
And the opposite signs of
second virial coefficient indicates that, 
the only difference in pressures between free
Fermi and Bose gases is the Fermi degeneracy pressure\cite{Vieferes}.
Also, note that all
the even virial coefficients 
are zero after $A_2$. Both of these features are explained later, more carefully.
But it is reported in ref.\cite{turza}  the equivalence found in 
Two-dimensional quantum gases is lost when they are trapped with potential.
So, let us now turn our attention towards the virial coefficients of the trapped system.
\subsection{Effect of trapping potential on virial coefficients}
To see the effect of different trapping potential
in  the various dimension we introduce two tables presenting virial coefficients
of Bose (table 1) and Fermi gases (table 2) below.
In this case we have considered symmetric potential, i.e. $n_1=n_2=..=n_d=n$.
It is clear from the tables that
the trapping potential  greatly affects
the virial coefficients.\\ \\

\begin{table}[h!]\footnotesize
\centering

\newcommand{\tvi}{\vrule height 10pt depth 10pt width -3pt} 
\newcolumntype{x}[1]{>{\hfil$\displaystyle} p{#1} <{$\hfil}} 

\begin{tabular}{|cc|ccccccc|}

\cline{3-9}
\multicolumn{2}{c|}{} & $v_1$ & $v_2$ & $v_3$ & $v_4$ & $v_5$ & $v_6$ & $v_7$
\\
\cline{3-9}

\hline
\tvi & {$n=2$} 
\tvi & {$1$}
\tvi & {$-2.50 \times 10^{-1}$} 
\tvi & {$2.78 \times 10^{-2}$} 
\tvi & {$0$} 
\tvi & {$-2.78 \times 10^{-4}$}
\tvi & {$0$} 
\tvi & {$4.72 \times 10^{-6}$}
\\

\tvi & {$n=5$} 
\tvi & {$1$} 
\tvi & {$-3.08 \times 10^{-1}$} 
\tvi & {$6.995 \times 10^{-2}$} 
\tvi & {$-1.15 \times 10^{-2}$} 
\tvi & {$9.99 \times 10^{-4}$}
\tvi & {$1.10 \times 10^{-4}$} 
\tvi & {$-6.17 \times 10^{-5}$}   
\\

\multirow{0}{*}{$d=1$}

\tvi & {$n=7$}
\tvi & {$1$}  
\tvi & {$-3.20 \times 10^{-1}$} 
\tvi & {$8.12 \times 10^{-2}$} 
\tvi & {$-1.62 \times 10^{-2}$} 
\tvi & {$2.23 \times 10^{-3}$} 
\tvi & {$-6.19 \times 10^{-5}$}
\tvi & {$-7.59 \times 10^{-5}$}
\\ 

\tvi & {$n=10$}
\tvi & {$1$}  
\tvi & {$-3.30 \times 10^{-1}$} 
\tvi & {$9.04 \times 10^{-2}$} 
\tvi & {$-2.06 \times 10^{-2}$} 
\tvi & {$3.61 \times 10^{-3}$} 
\tvi & {$-3.46 \times 10^{-4}$}
\tvi & {$-5.60 \times 10^{-5}$}
\\

\tvi & {$n=\infty$}
\tvi & {$1$}  
\tvi & {$-3.54 \times 10^{-1}$} 
\tvi & {$1.15 \times 10^{-1}$} 
\tvi & {$-3.41 \times 10^{-2}$} 
\tvi & {$9.01 \times 10^{-3}$} 
\tvi & {$-2.0 \times 10^{-3}$}
\tvi & {$3.12 \times 10^{-4}$}
\\
\hline

\tvi & {$n=2$} 
\tvi & {$1$}
\tvi & {$-1.25 \times 10^{-1}$} 
\tvi & {$-1.16 \times 10^{-2}$} 
\tvi & {$-2.60 \times 10^{-3}$} 
\tvi & {$-7.98 \times 10^{-4}$}
\tvi & {$-2.87 \times 10^{-4}$} 
\tvi & {$-1.14 \times 10^{-4}$}
\\

\tvi & {$n=5$} 
\tvi & {$1$} 
\tvi & {$-1.90 \times 10^{-1}$} 
\tvi & {$3.89 \times 10^{-4}$} 
\tvi & {$4.66 \times 10^{-4}$} 
\tvi & {$7.91 \times 10^{-5}$}
\tvi & {$1.24 \times 10^{-5}$} 
\tvi & {$2.07 \times 10^{-6}$}   
\\

\multirow{0}{*}{$d=2$}

\tvi & {$n=7$}
\tvi & {$1$}  
\tvi & {$-2.05 \times 10^{-1}$} 
\tvi & {$5.88 \times 10^{-3}$} 
\tvi & {$9.76 \times 10^{-4}$} 
\tvi & {$8.75 \times 10^{-5}$} 
\tvi & {$4.64 \times 10^{-6}$}
\tvi & {$-7.83 \times 10^{-8}$}
\\ 

\tvi & {$n=10$}
\tvi & {$1$}  
\tvi & {$-2.18 \times 10^{-1}$} 
\tvi & {$1.11 \times 10^{-2}$} 
\tvi & {$1.14 \times 10^{-3}$} 
\tvi & {$2.43 \times 10^{-5}$} 
\tvi & {$-8.83 \times 10^{-6}$}
\tvi & {$-1.36 \times 10^{-6}$}
\\

\tvi & {$n=\infty$}
\tvi & {$1$}  
\tvi & {$-2.50 \times 10^{-1}$} 
\tvi & {$2.78 \times 10^{-2}$} 
\tvi & {$0$} 
\tvi & {$-2.78 \times 10^{-4}$}
\tvi & {$0$} 
\tvi & {$4.72 \times 10^{-6}$}
\\
\hline

\tvi & {$n=2$} 
\tvi & {$1$}
\tvi & {$-6.25 \times 10^{-2}$} 
\tvi & {$-9.07 \times 10^{-3}$} 
\tvi & {$-2.71 \times 10^{-3}$} 
\tvi & {$-1.08 \times 10^{-3}$}
\tvi & {$-5.03 \times 10^{-4}$} 
\tvi & {$-2.59 \times 10^{-4}$}
\\

\tvi & {$n=5$} 
\tvi & {$1$} 
\tvi & {$-1.17 \times 10^{-1}$} 
\tvi & {$-1.20 \times 10^{-2}$} 
\tvi & {$-2.87 \times 10^{-3}$} 
\tvi & {$-9.32 \times 10^{-4}$}
\tvi & {$-3.54 \times 10^{-4}$} 
\tvi & {$-1.48 \times 10^{-4}$}   
\\

\multirow{0}{*}{$d=3$}

\tvi & {$n=7$}
\tvi & {$1$}  
\tvi & {$-1.31 \times 10^{-1}$} 
\tvi & {$-1.11 \times 10^{-2}$} 
\tvi & {$-2.36 \times 10^{-3}$} 
\tvi & {$-6.89 \times 10^{-4}$} 
\tvi & {$-2.37 \times 10^{-4}$}
\tvi & {$-8.97 \times 10^{-5}$}
\\ 

\tvi & {$n=10$}
\tvi & {$1$}  
\tvi & {$-1.44 \times 10^{-1}$} 
\tvi & {$-9.81 \times 10^{-3}$} 
\tvi & {$-1.81 \times 10^{-3}$} 
\tvi & {$-4.73 \times 10^{-4}$} 
\tvi & {$-1.47 \times 10^{-4}$}
\tvi & {$-5.03 \times 10^{-5}$}
\\

\tvi & {$n=\infty$}
\tvi & {$1$}  
\tvi & {$-1.77 \times 10^{-1}$} 
\tvi & {$-3.30 \times 10^{-3}$} 
\tvi & {$-1.11 \times 10^{-4}$} 
\tvi & {$-3.54 \times 10^{-6}$}
\tvi & {$-8.39 \times 10^{-8}$} 
\tvi & {$-3.66 \times 10^{-10}$}
\\
\hline

\end{tabular}
\caption{Virial coefficients of ideal Bose gas trapped under  power law potential  in d-dimension.}
\end{table}

\begin{table}[h!]
\footnotesize
\centering

\newcommand{\tvi}{\vrule height 10pt depth 10pt width -3pt} 
\newcolumntype{x}[1]{>{\hfil$\displaystyle} p{#1} <{$\hfil}} 

\begin{tabular}{|cc|ccccccc|}

\cline{3-9}
\multicolumn{2}{c|}{} & $v_1$ & $v_2$ & $v_3$ & $v_4$ & $v_5$ & $v_6$ & $v_7$
\\
\cline{3-9}

\hline
\tvi & {$n=2$} 
\tvi & {$1$}
\tvi & {$2.50 \times 10^{-1}$} 
\tvi & {$2.78 \times 10^{-2}$} 
\tvi & {$0$} 
\tvi & {$-2.78 \times 10^{-4}$}
\tvi & {$0$} 
\tvi & {$4.72 \times 10^{-6}$}
\\

\tvi & {$n=5$} 
\tvi & {$1$} 
\tvi & {$3.08 \times 10^{-1}$} 
\tvi & {$6.995 \times 10^{-2}$} 
\tvi & {$1.15 \times 10^{-2}$} 
\tvi & {$9.99 \times 10^{-4}$}
\tvi & {$-1.10 \times 10^{-4}$} 
\tvi & {$-6.17 \times 10^{-5}$}   
\\

\multirow{0}{*}{$d=1$}

\tvi & {$n=7$}
\tvi & {$1$}  
\tvi & {$3.20 \times 10^{-1}$} 
\tvi & {$8.12 \times 10^{-2}$} 
\tvi & {$1.62 \times 10^{-2}$} 
\tvi & {$2.23 \times 10^{-3}$} 
\tvi & {$6.19 \times 10^{-5}$}
\tvi & {$-7.59 \times 10^{-5}$}
\\ 

\tvi & {$n=10$}
\tvi & {$1$}  
\tvi & {$3.30 \times 10^{-1}$} 
\tvi & {$9.04 \times 10^{-2}$} 
\tvi & {$2.06 \times 10^{-2}$} 
\tvi & {$3.61 \times 10^{-3}$} 
\tvi & {$3.46 \times 10^{-4}$}
\tvi & {$-5.60 \times 10^{-5}$}
\\

\tvi & {$n=\infty$}
\tvi & {$1$}  
\tvi & {$3.54 \times 10^{-1}$} 
\tvi & {$1.15 \times 10^{-1}$} 
\tvi & {$3.41 \times 10^{-2}$} 
\tvi & {$9.01 \times 10^{-3}$} 
\tvi & {$2.0 \times 10^{-3}$}
\tvi & {$3.12 \times 10^{-4}$}
\\
\hline

\tvi & {$n=2$} 
\tvi & {$1$}
\tvi & {$1.25 \times 10^{-1}$} 
\tvi & {$-1.16 \times 10^{-2}$} 
\tvi & {$2.60 \times 10^{-3}$} 
\tvi & {$-7.98 \times 10^{-4}$}
\tvi & {$2.87 \times 10^{-4}$} 
\tvi & {$-1.14 \times 10^{-4}$}
\\

\tvi & {$n=5$} 
\tvi & {$1$} 
\tvi & {$1.90 \times 10^{-1}$} 
\tvi & {$3.89 \times 10^{-4}$} 
\tvi & {$-4.66 \times 10^{-4}$} 
\tvi & {$7.91 \times 10^{-5}$}
\tvi & {$-1.24 \times 10^{-5}$} 
\tvi & {$2.07 \times 10^{-6}$}   
\\

\multirow{0}{*}{$d=2$}

\tvi & {$n=7$}
\tvi & {$1$}  
\tvi & {$2.05 \times 10^{-1}$} 
\tvi & {$5.88 \times 10^{-3}$} 
\tvi & {$-9.76 \times 10^{-4}$} 
\tvi & {$8.75 \times 10^{-5}$} 
\tvi & {$-4.64 \times 10^{-6}$}
\tvi & {$-7.83 \times 10^{-8}$}
\\ 

\tvi & {$n=10$}
\tvi & {$1$}  
\tvi & {$2.18 \times 10^{-1}$} 
\tvi & {$1.11 \times 10^{-2}$} 
\tvi & {$-1.14 \times 10^{-3}$} 
\tvi & {$2.43 \times 10^{-5}$} 
\tvi & {$8.83 \times 10^{-6}$}
\tvi & {$-1.36 \times 10^{-6}$}
\\

\tvi & {$n=\infty$}
\tvi & {$1$}  
\tvi & {$2.50 \times 10^{-1}$} 
\tvi & {$2.78 \times 10^{-2}$} 
\tvi & {$0$} 
\tvi & {$-2.78 \times 10^{-4}$}
\tvi & {$0$} 
\tvi & {$4.72 \times 10^{-6}$}
\\
\hline

\tvi & {$n=2$} 
\tvi & {$1$}
\tvi & {$6.25 \times 10^{-2}$} 
\tvi & {$-9.07 \times 10^{-3}$} 
\tvi & {$2.71 \times 10^{-3}$} 
\tvi & {$-1.08 \times 10^{-3}$}
\tvi & {$5.03 \times 10^{-4}$} 
\tvi & {$-2.59 \times 10^{-4}$}
\\

\tvi & {$n=5$} 
\tvi & {$1$} 
\tvi & {$1.17 \times 10^{-1}$} 
\tvi & {$-1.20 \times 10^{-2}$} 
\tvi & {$2.87 \times 10^{-3}$} 
\tvi & {$-9.32 \times 10^{-4}$}
\tvi & {$3.54 \times 10^{-4}$} 
\tvi & {$-1.48 \times 10^{-4}$}   
\\

\multirow{0}{*}{$d=3$}

\tvi & {$n=7$}
\tvi & {$1$}  
\tvi & {$1.31 \times 10^{-1}$} 
\tvi & {$-1.11 \times 10^{-2}$} 
\tvi & {$2.36 \times 10^{-3}$} 
\tvi & {$-6.89 \times 10^{-4}$} 
\tvi & {$2.37 \times 10^{-4}$}
\tvi & {$-8.97 \times 10^{-5}$}
\\ 

\tvi & {$n=10$}
\tvi & {$1$}  
\tvi & {$1.44 \times 10^{-1}$} 
\tvi & {$-9.81 \times 10^{-3}$} 
\tvi & {$1.81 \times 10^{-3}$} 
\tvi & {$-4.73 \times 10^{-4}$} 
\tvi & {$1.47 \times 10^{-4}$}
\tvi & {$-5.03 \times 10^{-5}$}
\\

\tvi & {$n=\infty$}
\tvi & {$1$}  
\tvi & {$1.77 \times 10^{-1}$} 
\tvi & {$-3.30 \times 10^{-3}$} 
\tvi & {$1.11 \times 10^{-4}$} 
\tvi & {$-3.54 \times 10^{-6}$}
\tvi & {$8.39 \times 10^{-8}$} 
\tvi & {$-3.66 \times 10^{-10}$}
\\
\hline
\end{tabular}
\caption{Virial coefficients of ideal Fermi gas trapped  under  power law potential d-dimension}
\end{table}
More importantly, it is seen in the tables that the virial coefficients of
 one dimensional harmonically
trapped Bose gases are equal to 
the  one dimensional harmonically
trapped Fermi gases,
except the second virial coefficients 
for which the signs are opposite.
This outstanding property, ealier seen in two-dimensional free quantum gases,
indicates
the equivalence reported by Mehedi et.al.\cite{turza}.
Another intriguing behavior is the appearance
of even virial coefficients 
being zero while there is an equivalence
in Bose and Fermi gases are seen in this case, which is also found in two-dimensional free quantum gases only.
And finally we point out the most remarkable property feature in the table, which is 
the virial coefficients of one-dimensional harmonically trapped quantum gases
are being equal to the virial coefficients of two dimensional free system (both bosons and fermions).
This phenomenon suggests high-temperature behavior of  two-dimensional free quantum gases (Bose and Fermi) is similar to
one dimensional harmonically trapped quantum gases (Bose and Fermi).\\\\
\subsubsection{Equivalence of harmonically trapped quantum gases in d=1}
It is very important to point out in both the cases where this equivalence of Bose and 
Fermi gases are noticed\cite{turza}
takes the value $\chi=1$.
Now, focusing over the harmonically trapped quantum gases in one dimension, we find the density of such quantum gases
\begin{eqnarray}
 \rho=sgn(\sigma)\frac{1}{\lambda'} Li_1(\sigma)=\frac{1}{\lambda'} log(1\pm z)
\end{eqnarray}
where, the upper sign is for fermions and the other is for boson.
From this we can explicitly write the fugacity, which is
\begin{equation}
 z=\mp 1\pm e^{\pm \rho\lambda'}
\end{equation}
The pressure equation leads us to
\begin{equation}
 \beta P=\mp\frac{1}{\lambda'} Li_2(1\pm z)
\end{equation}
Now using the techniques of partial derivative
$(\frac{\partial P}{\partial \rho})_{_T}=(\frac{\partial P}{\partial \mu})_{_T}(\frac{\partial \mu}{\partial \rho})_{_T}$, we get from the above two equations,
\begin{eqnarray}
\bigg(\frac{\partial P}{\partial \rho} \bigg)_T =\mp \frac{\lambda'}{\beta}\frac{1}{e^{\mp \rho\lambda'}-1}
\end{eqnarray}
Now the equation of state is thus given by,
\begin{equation}
 \beta P= \mp  \int_0 ^\rho \frac {\gamma \lambda'}{e^{\mp\rho\lambda'}-1} d\gamma
\end{equation}
Notably this functional form can be expressed in terms of
Bernoulli numbers\cite{Vieferes},
\begin{equation}
 \frac{p}{e^p-1}= \sum_{n=0} ^{\infty} B_n \frac{p^n}{n!}
\end{equation}
where, $0<|p|<2\pi$. $B_n$
are known as Bernoulli numbers and be defined as\cite{new},
\begin{eqnarray}
&& B_n = \left\{    \begin{array}{lr}
     0 &,  \text{n=0}\\
     -\frac{1}{2} &,  \text{n=1}\\
     (-1)^{\frac{n}{2}-1}  \frac{2(2n)!}{(2\pi)^{2n}}\zeta(2n) &,  \text{even n}\\
      0 &,  \text{odd n}
     \end{array}
   \right.
\end{eqnarray}\\
Again rewriting Eq. (21), choosing $d=1$ and $n=2$,
\begin{eqnarray}
 \beta P=\sum_{k=1} ^{\infty}A_k \rho ^k \lambda'^{(k-1)} 
\end{eqnarray}
Using the above equations we conclude virial coefficients in this case,
\begin{equation}
A_l=(\mp)^{l-1}\frac{B_{l-1}}{l!} 
\end{equation}
Incidentally this is the same result obtained in two dimensional free quantum
gas as well\cite{Vieferes}. Now as odd Bernoulli numbers $B_n$ are zero except $(n=1)$
we find out even virial coefficients $A_n$ will be zero for both Bose and Fermi gases, except $(n=2)$.
As it is seen from above beside the second virial coefficient, all the virial coefficients do become the same 
when there is an equivalence, let us focus on this phenomena in a more detail.
The pressure for one dimensional harmonically trapped quantum gases from equation (38),
\begin{eqnarray}
 \beta P =  \rho
 \pm \frac{1}{4}\rho ^2\lambda' + 
 \frac{1}{36}\rho^3 \lambda'^2-
\frac{1}{3600} \rho^5 \lambda'^4
+ \frac{1}{211680}\rho^7 \lambda'^6 
\end{eqnarray}
So, the difference in
pressure of Bose and Fermi gas in this case is,
\begin{eqnarray}
 P_F-P_B= \frac{1}{2\beta} \rho^2 \lambda'
  = \frac{1}{2\beta} \frac{N}{V'} \lambda'
\end{eqnarray}
Now, with harmonic trapping potential $V'\varpropto T^{1/2}$ and 
$\lambda' \varpropto T^{-\frac{1}{2}}$, the right hand side of the above equation is a temperature independent quantity.
As it turns out it is nothing but the ground state pressure also
known as degeneracy pressure in Fermi gases\cite{turza}.
similar situation is also observed in ref. 
\cite{Vieferes} in the case of two dimensional free quantum gases. 
The reason   of such phenomena is due to Landen relation of dilog functions\cite{lee},
\begin{eqnarray}
 Li_2(x_1)=-Li_2(x_2)-\frac{1}{2} [Li_1(x)]^2
\end{eqnarray}
where, $x_1$ and $x_2$ are related as $x_2=\frac{x_1}{1-x_1}$, known as Euler transformation\cite{lee1}.
As fugacities of Bose and Fermi gas can be connected 
as a Euler transformation 
eq. (10) and (11) dictates the relation described by eq. (43)

\section{Conclusion}
The virial coefficients of both types of ideal quantum gases trapped under generic power
law potentials
are calculated
from unified statistical thermodynamics. The general results of this paper 
coincide with
the known results\cite{pathria,Vieferes} in $d=2$ and $d=3$
with appropriate  choice
of $n_i$. 
The equivalence\cite{lee} of two-dimensional ideal free
Bose
and Fermi gases revealed a remarkable property\cite{Vieferes}
that their virial coefficient are same 
(except the second virial coefficient,
where the sign is different).
We further showed that,
the recently established
equivalence (in $d=1$)
between the harmonically trapped ideal
Bose and Fermi gases 
also maintains this
property.  Hence it can be concluded that, in both of the cases, where equivalence relation 
can be established between Bose and Fermi gas, bosons and fermions do 
carry identical virial coefficients.
Also from the table, one can see 
that the 
 virial coefficients of two-dimensional free quantum gases are identical to 
 the virial coefficients of one-dimensional harmonically trapped quantum gases.
This interesting result suggests that the high-temperature behavior of bosons and fermions
in these two types of systems should be indistinguishable.
Lastly, since the calculation in this paper is done in thermodynamic limit, 
the virial coefficients
are yet to be done beyond the thermodynamic limit.
We are currently doing this using Yukolov's
semiclassical 
approximation\cite{Yukalov}. 
It will be equally intriguing to examine 
the behavior of virial coefficients for relativistic quantum gases 
by taking into account the effect of antiparticles.
\section{Acknowledgement}
MMF would like to thank Md. Alamgir Al
Faruqui for his cordial hospitality during MMF's stay in London. Special thanks to Laura Alejandra Moya Camacho to help MMF to present the manuscript.


\begin{thebibliography}{1}


 \bibitem{bose} S. N. Bose, Z. Physik, 26, 178 (1924).
\bibitem{einstein} A. Einstein, Berl. Ber 22, 261 (1924).
 \bibitem{ziff} R. M. Ziff, G. E Uhlenbeck, M. Kac, Phys. Reports 32 169 (1977).
\bibitem{pathria} R. K. Pathria, Statistical Mechanics, Elsevier, 2004.
\bibitem{howard}H. E. Haber and H. A. Weldon, Phys. Rev. Lett.  46 (1981) 
\bibitem{may} R. H. May, Phys. Rev., A1515, 135, (1964).
   \bibitem{r} J.E. Robinson, Phys Rev. E 83, 678 (1951).
   \bibitem{lieb1} E. H. Lieb and W. Liniger, Phys. Rev. 130, 1605 (1964).
   \bibitem{lieb2} E. H. Lieb, Phys. Rev. 130, 1606 (1963).
   \bibitem{fisher} D. S. Fisher and P. C. Hohenberg, Phys. Rev. B 37, 4936 (1988).  
   \bibitem{beckmann} R. Beckmann, F. Karch, D.E. Miller, Phys. Rev. Lett., 43 1277 (1979).
\bibitem{b1}
Bender, C. M., Boettcher, S., and Lipatov L. (1992). Phys. Rev. D 46, 5557.
\bibitem{b2}
Bender, C. M., and Boettcher, S. (1994). Phys. Rev. D 51, 1875.
\bibitem{b3}
Bender, C. M., and Boettcher, S., and Mead L. R. (1994). J. Math. Phys. 35, 368.
\bibitem{b4}
Bender, C. M., and Milton, K. A. (1994). Phys. Rev. D 50, 6547.
\bibitem{b5}
Grosche, C., and Steiner, F. (1995). J. Math. Phys. 36, 2354.
   \bibitem{lee}  M. H. Lee, Phys. Rev. E, 55, 1518-1520 (1997).
\bibitem{lee1} M. H. Lee,  Acta Phys. Polonica, 40, 1279-1301 (2009)
\bibitem{Vieferes}S. Viefers, F. Ravndal, T. Haugset American Journal of Physics 63, 369 (1995); doi: 10.1119/1.17922
 
 \bibitem{Bradley} C. C. Bradley, C. A. Sackett, J. J. Tollett and R. G. Hulet, Phys. Rev. Lett. 75, 1687, 1995.
\bibitem{anderson} M. H. Anderson, J. R. Esher, M. R. Mathews, C. E. Wieman and E. A. Cornell, Science 269, 195, 1995.
 \bibitem{davis} K. B. Davis, M. O. Mewes, M. R. Andrew, N. J. Van Druten, D. S. Durfee, D. M. Kurn and W. Ketterle, 
 Phys. Rev. Lett., 1687, 75, 1995.
\bibitem{DeMarco} DeMarco B and Jin D S 1999 Science 285 1703.

 \bibitem{s} S Biswas, J Mitra, S Bhattacharyya, J. Stat. Mech. P03013, 2015.
\bibitem{sala} Luca Salasnich, J. Math. Phys 41, 8016 (2000).
\bibitem{yan}Z. Yan, Phys. Rev. A 59, 1999. 
\bibitem{yan1}Z. Yan, Phys. Rev. A 61, 2000.
\bibitem{yan2}Z. Yan, Mingzhe Li, L Chen, C. Chen and J. Chen,
J. Phys. A: Math. Gen. 32 (1999) 4069–4078.
\bibitem{yan3}Z. Yan,   Eur. J. Phys. 21 625,  2000. 
\bibitem{m1}M M Faruk and G. M. Bhuiyan, Acta Physica Polonica B, Vol. 46 (2015)
\bibitem{champu}M M Faruk, Acta Physica Polonica B, Vol. 46 (2015)
\bibitem{Bagnato} Bagnato, V., and Kleppner, D. (1991). Phys. Rev. A, 44(11), 7439.
\bibitem{Dai} Dai, W. S., and Xie, M. (2003). Phys. Rev. A, 67(2), 027601.
\bibitem{turza}M M Faruk, J Stat Phys, DOI 10.1007/s10955-015-1344-4.
\bibitem{igu} Kazumoto Iguchi, Mod. Phys. Lett. B, 11, 765 (1997)

\bibitem{new} Abramowitz, M. and Stegun, I. A. (Eds.).
Handbook of Mathematical Functions with Formulas, Graphs, and Mathematical Tables, 9th printing. New York: Dover, pp. 804-806, 1972. 
   \bibitem{Yukalov}V. I. Yukalov,   Phys Rev. A 72, 033608 (2005). 
\end{thebibliography}
 \end{document}